# Water vapour effects in mass measurement

N. Khélifa

LNE-INM : Conservatoire national des arts et métiers
61, rue du Landy, 93210, La Plaine Saint Denis, France
Email: khelifa@cnam.fr

***Abstract.*** *Water vapour inside the mass comparator enclosure is a critical parameter. In fact, fluctuations of this parameter during mass weighing can lead to errors in the determination of an unknown mass. To control that, a proposal method is given and tested. Preliminary results of our observation of water vapour sorption and desorption processes from walls and mass standard are reported.*

## 1. Introduction

Water vapour contained in ambient air influences measurement in many field of science. In mass metrology, it is necessary to correctly evaluate air moisture and its changes during mass comparisons. In practice, the air density $\rho_{air}$ [kg m$^{-3}$] is deduced, by using the so called "CIPM- 1981/91 formula" for air density [1], from the measured parameters of ambient air.

$$\rho_{air} = \frac{p M_a}{Z R T}\left[1 - x_v\left(1-\frac{M_v}{M_a}\right)\right] + \varepsilon_f$$

In this formula, $M_a$ and $M_v$ are the molar mass [kg mol$^{-1}$] of respectively dry air and water, Z the compressibility factor, R the constant of ideal gases [J mol$^{-1}$ K$^{-1}$], p the atmospheric pressure [Pa], $x_v$ the molar fraction of water vapor and $\varepsilon_f = \pm 0{,}7 \times 10^{-4}$ kg m$^{-3}$ the fitting error of the formula.

The combined relative standard uncertainty of air density, for normal atmospheric conditions, is at the level of $1 \times 10^{-4}$ kg m$^{-3}$, when uncertainties for p, T, $M_a$ and $x_v$ are included [1]. Recently, determination of air density from the results of artefacts comparisons in ambient air and in vacuum, show a significant difference against the value evaluated by the classical method (CIPM- 1981/91 formula). This discrepancy is of the order of the fitting error of the CIPM formula, i.e. $0{,}7 \times 10^{-4}$ kg m$^{-3}$ [2, 3]. This observation needs further studies, particularly concerning the existing air density gradients inside the mass comparator chamber. In fact, sorption and desorption of water vapor by the mass standards and the walls of the chamber, during the weighing procedure induce fluctuations of air inside the enclosure.

## 2. Air moisture measurement inside the enclosure

In practice the molar fraction of water vapor is calculated from the measurement of dew point temperature $t_d$ or from the relative humidity HR. In order to minimize air perturbations inside the chamber of the mass balance, the temperature of the dew point is measured only a long time before the start and at the end of a succession of weighing. During mass weighting, which takes several hours, sometime a capacitive hygrometer is used to evaluate the drift in air humidity. Here we use a method, based on molecular absorption in the near infrared, developed and assigned to follow the small changes of water vapour. The near-infrared laser diode device is described in a previous reports [4, 5]. Here, we use this sensor to confirm the indiscreet behaviour of the dew-point hygrometer and to observe water vapor sorption and desorption.

A schematic diagram of the experimental setup is represented by figure 1. The systems consist of a vacuum housing and an oil free pump associated to a turbo molecular pump. The working pressure range is between 1 Pa and standard atmosphere ($10^5$ Pa). The pressure inside the enclosure (vacuum pump shut off) goes from 1 Pa to 1800 Pa in about 120 minutes. Initial sorption and desorption processes within the vacuum is masked by the global leak. With the "made home" optical hygrometer, air moisture is monitored when air is progressively evacuated from the housing.

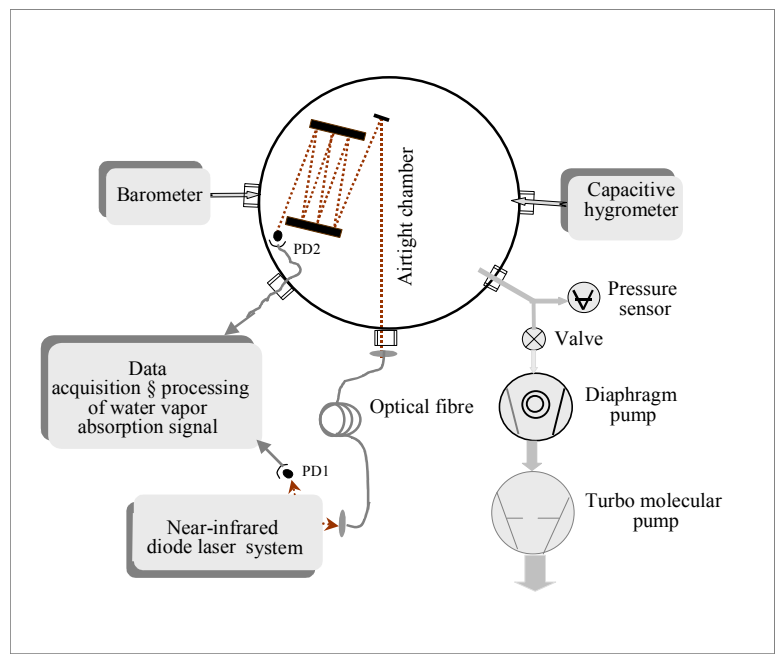

Fig. 1. Schematic diagram of experiment

The water vapor absorption is converted onto partial pressure. For that, the absorption signal is adjusted to the mean value of relative humidity measured, with a calibrated capacitive hygrometer, at the start of measurement.

Figure 2 shows the changes of water vapor inside the enclosure during time. The air environmental parameters, are $\overline{p} = 101532\ Pa$, $t = 21{,}30\ °C$ and a relative humidity HR = 52,8 %. The results show the perturbations induced by the dew point hygrometer and confirm the superiority of the optical device compared to the capacitive in terms of sensitivity.

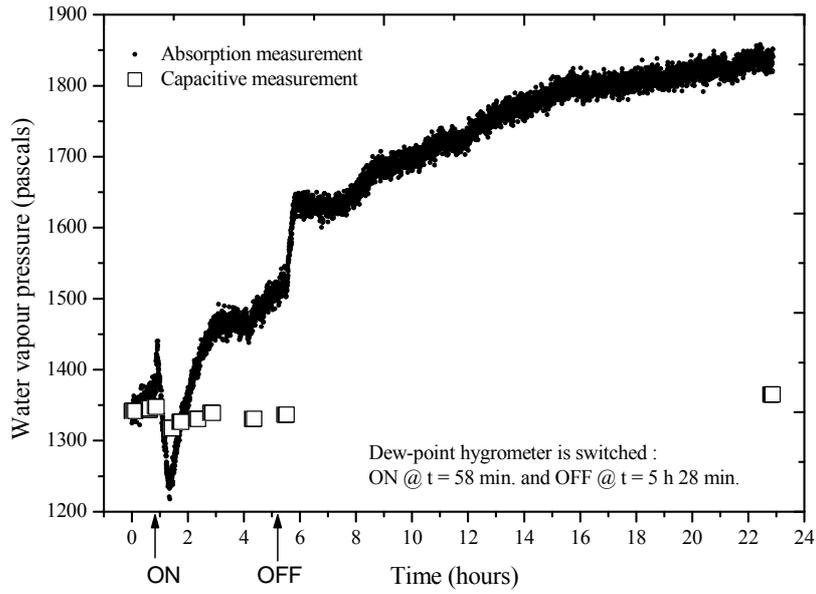

Fig. 2. Evolution of relative humidity within total air pressure inside the housing.

The changes of water vapor, when total pressure of moist air inside the chamber is progressively diminished by pumping are given in figure 3. As we can see, below a air pressure of about 80000 Pa, the water vapor decrease very slowly.

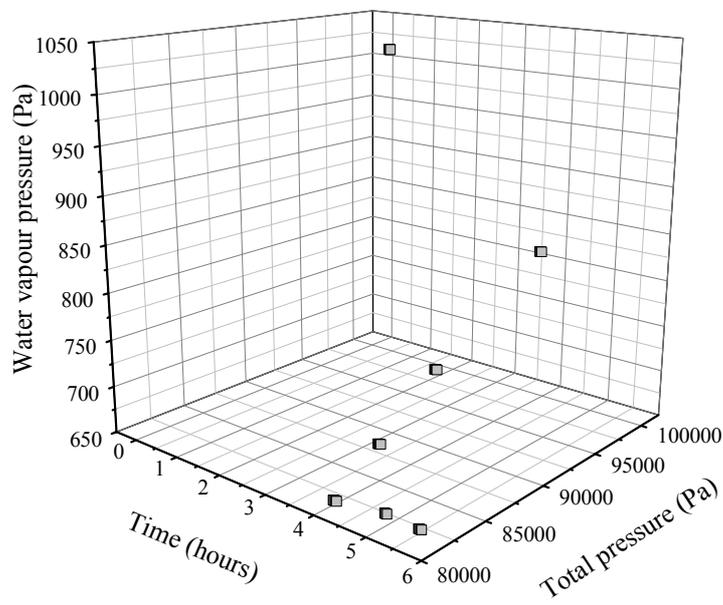

Fig. 3. Variation of water vapor pressure as a function of total air pressure

## 3. Conclusion

In vacuum comparisons, supplementary correction must be made due to sorption process. The value of adsorbed mass per surface area depends on mass standard (material allow, surface roughness, …). In fact, a monolayer of adsorbed water vapor on the surface of a stainless steel kilogram standard corresponds to a mass variation of about $2\,\mu g$. In mass comparisons, the effects of buoyancy in air and the adsorption of air molecules onto the surface of weight need to be considered.